\documentclass[prb,twocolumn,aps,showpacs,fixfloats]{revtex4}
\usepackage{graphicx}
\usepackage{bm}
\usepackage{amsmath,amssymb}
\usepackage{subfigure}
\usepackage{float}
\usepackage{latexsym}
\usepackage{color}
\usepackage{enumerate}
\usepackage{pdfpages}
\usepackage{tikz}
\usepackage{hyperref}
\usepackage{graphicx}
\usepackage{bm}
\usepackage{amssymb} 
\usepackage{amsmath}
\usepackage{subfigure}
\newcommand{\s}{\scriptscriptstyle}
\usepackage{relsize}
\begin{document}


\title{Renormalization of the 3D exciton spectrum by the disorder }

\author{R. E. Putnam, Jr. and M. E. Raikh}

\affiliation{ Department of Physics and
Astronomy, University of Utah, Salt Lake City, UT 84112}

\begin{abstract}
Effect of short-range disorder on the excited states of the exciton
is studied. Disorder causes an obvious effect of broadening. Microscopically,
an exciton, as an entity, is scattered by the large-scale disorder fluctuations.
Much less trivial is that short-scale fluctuations, with a period of the order of
the Bohr radius, cause a well-defined down-shift of the exciton levels. We demonstrate
that this shift exceeds the broadening parametrically and study the dependence of this
shift on the orbital number. Difference of the shifts for neighboring levels leads to
effective renormalization of the Bohr energy. Most remarkable effect is the disorder-induced splitting of S and P exciton levels. The splitting originates
from the fact that disorder lifts the accidental degeneracy of the hydrogen-like levels.
The draw an analogy between this splitting and the Lamb shift in quantum electrodynamics.
\end{abstract}

\maketitle

\section{Introduction}

Early measurements\cite{1952,1952',1955} of optical absorption in Cu$_2$O crystal films
have revealed a series of lines below the main absorption edge. Remarkably, the positions of these lines could be fitted with a hydrogen-like spectrum with high accuracy.
For this reason, the series were identified with ground and excited states of the exciton.

Interest to highly excited (Rydberg) states of excitons has been
revived after the publication of the paper Ref.~\onlinecite{Experiment0}.
In this paper, the existence of Rydberg excitons in Cu$_2$O with principle quantum number up to $n=25$ was established on the basis of the high-resolution absorption measurements. Observability of the Rydberg states in Cu$_2$O is explained by the high value of the Bohr energy in this crystals, namely, $94$meV\cite{Experiment0}.

In subsequent studies, see e.g. Refs.
\onlinecite{Glazov,Experiment1,microcrystals,linewidth,Chaos,Scaling,Yakovlev,Disorder},
many peculiarities of Rydberg excitons were revealed via the evolution of
the absorption lines with electric field. In particular, the deviation of the
binding energies from $\frac{1}{n^2}$ dependence was related to non-parabolicity
of the underlying bandstructure. Concerning the shapes of the absorption lines, it was found to be Lorentzian\cite{Experiment0}
for $n<10$ and evolves towards Gaussian with increasing $n$. The widths of the  lines
decreased with $n$ as $n^{-3}$ and saturated for $n>10$. It was concluded that the origin of the low-$n$ broadening is radiative recombination, while the higher-$n$ broadening is due to crystal inhomogeneities.

The effect of disorder (inhomogeneities) on Rydberg excitons
is an issue of conceptual importance. Indeed, the excited levels of hydrogen
possess an accidental degeneracy: $n^2$ wave functions correspond to the level, $n$.
This degeneracy is lifted in {\em any} given disorder configuration.
Thus, inhomogeneous broadening, originating from variation of configurations, is accompanied by splitting, which is also a disorder-induced effect.  A nontrivial question arises: which of the two effects is dominant? This question is addressed in the present paper. We consider the simplest model of an exciton in parabolic bands and in the presence of a weak  white-noise disorder.

%
.


\begin{figure}
\includegraphics[width=80mm]{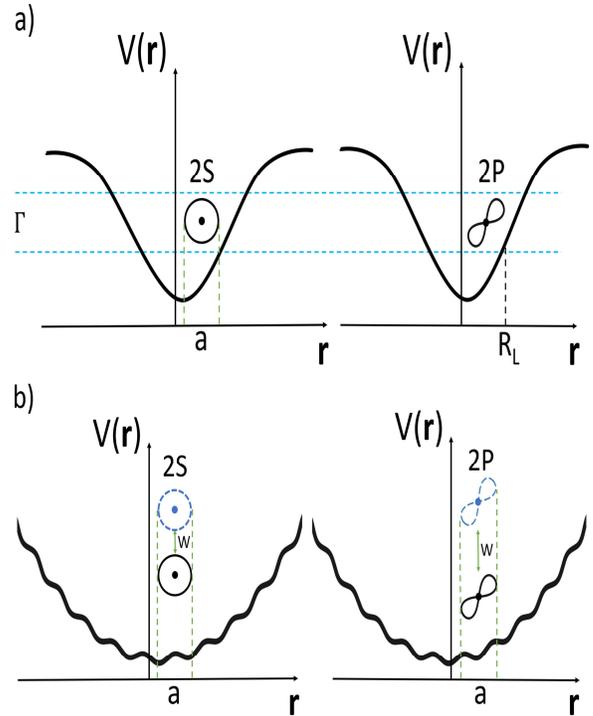}
\caption{(Color online) (a) Cartoon illustrating the disorder-induced
broadening of 2S and 2P excitons. Exciton, as a whole, is trapped in
a wide and shallow potential well of a size, $R_L$, given by Eq. (\ref{width}).
The well is created by a fluctuation.
The width, $\Gamma$, is of the order of the depth of the well
Eq. (\ref{Gamma}); (b) Small-scale
fluctuations with a size of the order of Bohr radius cause a down-shift
of the exciton level. The shift of 2P is bigger than the shift of 2S. }
\label{figure1}
\end{figure}

\section{Disorder-induced broadening}

For simplicity we assume that the disorder potential acts only on the
electron. Then the Hamiltonian of the exciton reads

\begin{equation}
\label{Hamiltonian}
 {\hat H} = \frac{{\hat p}_e^2}{2m_e} +\frac{{\hat p}_h^2}{2m_h} -
    \frac{e^2}{\kappa\vert{\bf r}_e -{\bf r}_h\vert}+V({\bf r}_e),
\end{equation}
where $m_e$ and $m_h$ are the masses of electron and hole, ${\hat p}_e$ and ${\hat p}_h$
are the corresponding momenta, and $\kappa$ is the dielectric constant. The magnitude
of the short-range disorder potential is specified by the correlation relation
\begin{equation}
\label{correlator1}
\langle V({\bf r})V({\bf r}')\rangle =\gamma \delta\left({\bf r}-{\bf r}'\right).
\end{equation}
Introducing the center of mass coordinate, ${\bf R}$, and the relative coordinate, ${\bf r}$, in a standard way, we rewrite the Hamiltonian Eq. (\ref{Hamiltonian}) in the form
\begin{equation}
\label{Hamiltonian'}
{\hat H}=-\frac{\hbar^2}{2M}\Delta_{\bf R} -\frac{\hbar^2}{2\mu}\Delta_{\bf r} -\frac{e^2}{\kappa r} +V\left({\bf R}+\frac{m_h}{M}{\bf r}  \right),
\end{equation}
where $M=m_e+m_h$ is the net mass, while $\mu=\frac{m_em_h}{M}$ is the reduced mass of the exciton. To find the width of the exciton line qualitatively, we first assume that characteristic $R=R_L$ is much bigger than the Bohr radius
\begin{equation}
\label{radius}
a=\frac{\hbar^2\kappa}{\mu e^2}.
\end{equation}
We will later inspect this assumption.

Characteristic magnitude $V=V_L$ of the random potential
corresponding to the spatial scale, $R_L$,
can be found from the relation Eq.~(\ref{correlator1})
\begin{equation}
\label{estimate}
V_L^2R_L^3=\gamma.
\end{equation}
An estimate for the width of the exciton level comes from the uncertainty principle,
\begin{equation}
\label{estimate'}
\Big(\frac{\hbar^2}{MV_L}\Big)^{1/2}=R_L,
\end{equation}
stating that the de-Broglie wavelength corresponding to the energy $V_L$ is of the order of $R_L$.
Solving the system Eqs. (\ref{estimate}), (\ref{estimate'}, we find
\begin{equation}
\label{width}
R_L =\frac{\hbar^4}{M^2\gamma}.
\end{equation}
We can now go back to the above assumption $R_L\gg a$. Using Eq. (\ref{width}), this assumption
can be cast in the form of the condition
\begin{equation}
\label{checking}
\gamma \ll \frac{\hbar^4}{M^2a}.
\end{equation}
If this condition is met, the magnitude, $V_L$, should
be identified with the width of the exciton line
\begin{equation}
\label{Gamma}
\Gamma \sim V_L=\frac{\gamma^2M^3}{\hbar^6}.
\end{equation}
For the above picture to be consistent, one should ensure that
the exciton is not destroyed by the disorder.
This implies that the width, $\Gamma$, is much smaller than the
Bohr energy $E_B=\frac{\hbar^2}{2\mu a^2}$. From Eq. (\ref{Gamma})
the ratio $\Gamma/E_B$ can be expressed as
\begin{equation}
\label{ratio}
\frac{\Gamma}{E_B}=\frac{\mu}{M}\Big(\frac{\gamma M^2a}{\hbar^4}   \Big)^2.
\end{equation}
Comparing Eqs. (\ref{checking}) and (\ref{ratio}), we conclude
that the conditions $a\ll R_L$ and $\Gamma \ll E_B$ are equivalent
to each other.
Overall scenario of the disorder-induced broadening of the exciton line is illustrated in
Fig. \ref{figure1}. The exciton, as an entity, is trapped in a shallow fluctuation-induced potential well of a large size. The depth of the potential well is the exciton linewidth. Within this scenario,
the linewidth is independent of the Bohr radius.

The above reasoning was introduced in
Refs. \onlinecite{Alferov,Baranovskii,broadening}. In Ref. \onlinecite{broadening}
a numerical factor in the expression Eq. (\ref{Gamma}) was estimated to be $0.08$.

\section{Disorder-induced Shift}
While the long-range potential fluctuations with characteristic scale $R_L\gg a$ are responsible for inhomogeneous broadening of the exciton line, the short-range fluctuations with a scale $\sim a$ lead to another observable effect: disorder-induced shift of the exciton levels.\cite{review}
see Fig. \ref{figure1}.
We will illustrate this effect for the ground state of the exciton with a wave function
\begin{equation}
\label{1S}
\psi_{1S}({\bf r})=\frac{1}{\pi^{1/2}a^{3/2}}\exp\left(-\frac{r}{a}   \right).
\end{equation}
For this purpose, we will specify the form of the effective
random potential acting on the center of mass.

Searching for the solution of the Scrhr{\"o}dinger equation
${\hat H}\Psi({\bf R},{\bf r})=E\Psi({\bf R},{\bf r})$  in the form
\begin{equation}
\label{form}
\Psi({\bf R},{\bf r})=C_{1S}({\bf R})\psi_{1S}({\bf r}).
\end{equation}
Smallness of the effect of disorder allows one to neglect the admixture of
the higher states of the exciton to the wave function and obtain
a closed equation for $C_{1S}({\bf R})$
\begin{equation}
\label{closed}
-\frac{\hbar^2}{2M}\Delta_{\bf R}C_{1S}({\bf R})+V_{\s eff}({\bf R})
C_{1S}({\bf R})
=\left(E-E_B\right)C_{1S}({\bf R}),
\end{equation}
where the effective potential $V_{\s eff}({\bf R})$ is defined as
\begin{equation}
\label{effective}
V_{\s eff}({\bf R})=\int d{\bf r}~V\left({\bf R}+\frac{m_h}{M}{\bf r}  \right)\Big(\psi_{1S}({\bf r})\Big)^2.
\end{equation}
The structure  of the effective potential suggests that the value of $V_{\s eff}$
at some point, ${\bf R}$, is the result of interaction of electron
with the disorder potential in the course of orbiting around the center
of mass located at ${\bf R}$.

Correlation properties of the effective potential follow from the relation
Eq. (\ref{correlator1}) for the bare potential

\begin{widetext}
\begin{align}
\label{corr1}
&\langle V_{\s eff}({\bf R}_1)V_{\s eff}({\bf R}_2)\rangle
=\int d{\bf r}_1 d{\bf r}_2
\Big<V\Big({\bf R}_1 + \frac{m_h}{M}{\bf r}_1\Big)V\Big({\bf R}_2 + \frac{m_h}{M}{\bf r}_2\Big)\Big>
\psi_{1S}^2({\bf r}_1)\psi_{1S}^2({\bf r}_2)\nonumber\\
&=\gamma \int d{\bf r}_1 d{\bf r}_2~\delta \Big( {\bf R}_1 + \frac{m_h}{M}{\bf r}_1                    -{\bf R}_2 + \frac{m_h}{M}{\bf r}_2\Big) \psi_{1S}^2({\bf r}_1)\psi_{1S}^2({\bf r}_2)     .
\end{align}
\end{widetext}
It is now convenient to switch to the Fourier transform of the $\delta$-function, which
allows to reduce the number of integrations in Eq. (\ref{corr1}). One has
\begin{align}
\label{corr2}
&\langle V_{\s eff}({\bf R}_1)V_{\s eff}({\bf R}_2)\rangle
=\frac{\gamma}{(2\pi)^3}\int d{\bf k}\exp \Big\{i{\bf k}\Big({\bf R}_1-{\bf R}_2\Big) \Big\}\nonumber\\
&\times\Big\vert \int d{\bf r}~\psi_{1S}({\bf r})^2 \exp \Big\{i\frac{m_h}{M}{\bf k}{\bf r}     \Big\}                     \Big\vert^2.
\end{align}
From the form of the correlator, we conclude that characteristic values of $|{\bf k}|$ are of
the order of $a^{-1}$. Thus the correlator decays at distances $|{\bf R}_1-{\bf R}_2|>a$.

Using Eq. (\ref{corr2}) we can express the second-order correction to the  energy of
the exciton moving with momentum ${\bf p}$ caused by $V_{\s eff}({\bf R})$

\begin{equation}
\label{second-order}
\delta E_{\bf p}= \sum_{{\bf p}'} \frac{\Big\vert \{V_{\s eff}({\bf R})\}_{{\bf p},{\bf p}'}\Big\vert^2}{E_{\bf p}-E_{{\bf p}'}},
\end{equation}
where $\{V_{\s eff}({\bf R})\}_{{\bf p},{\bf p}'}$ is the matrix element of the
effective potential between the states ${\bf p}$ and ${\bf p}'$, while $E_{\bf p}$
is the dispersion law of the exciton
\begin{equation}
\label{spectrum}
E({\bf p})=-E_B +\frac{\hbar^2{\bf p}^2}{2M}.
\end{equation}
The pole at $E_{{\bf p}'}=E_{\bf p}$ is responsible for the broadening of the exciton line.
In order to recover the result Eq.~(\ref{Gamma}) within a numerical factor, we note that the
estimate for the left-hand side is $\Gamma$, while the pole in the right-hand side leads to the
imaginary part proportional to the density of states  at energy
$E_{\bf p}^{1/2}\sim \Gamma$. Since the density of the free exciton
states is proportional to $E_{\bf p}^{1/2}$, the right-hand side is proportional to $\Gamma^{1/2}$.
Then Eq. (\ref{second-order}) can be viewed as an equation for $\Gamma$ and yields the solution
Eq.~(\ref{Gamma}).

It is important to note that the imaginary part of the sum Eq. (\ref{second-order})
comes from the momenta $p \sim R_L^{-1}$, while the real part is accumulated from
much  bigger domain of momenta $p \sim a^{-1}$. From this we conclude
that the real part of the sum is a self-averaging quantity, which permits the replacement of
$\Big\vert \{V_{\s eff}({\bf R})\}_{{\bf p},{\bf p}'}\Big\vert^2$ by its average and neglect $E_{\bf p}$
compared to $E_{{\bf p}'}$ in the denominator of Eq. (\ref{second-order}).
As a result, Eq. (\ref{second-order}) turns into a down-shift of
the ground-state level of the exciton

\begin{equation}
\label{shift1S}
W_{1S}=-\frac{\gamma}{(2\pi)^3}\int \frac{d{\bf p}'}{E_{{\bf p}'}}
\Big\vert  \int d{\bf r}~\psi_{1S}({\bf r})^2 \exp \Big\{-i\frac{m_h}{M}\left({\bf p}'{\bf r}\right)     \Big\}                          \Big\vert^2.
\end{equation}
The remaining task is to evaluate the integrals in Eq.~(\ref{shift1S}),
which can be done analytically.
The spatial integral is taken using the relation
\begin{equation}
\label{relation}
\int d{\bf r}~\psi_{1S}^2\exp(i{\bf qr})=\frac{1}{\Big(1+\frac{q^2a^2}{4}\Big)^2}.
\end{equation}
Now the integral over ${\bf p}'$ can be made dimensionless and evaluated.
The result reads
\begin{equation}
\label{result}
W_{1S}=-\frac{\gamma M^{2}}{\pi^2\hbar^2m_h^a}
\int\limits_0^{\infty}\frac{dz}{\Big(   1 + \frac{z^2}{4}\Big)^4}
=-\frac{5\gamma M^{2}}{16\pi\hbar^2m_ha}.
\end{equation}
Recall now, that we have restricted the calculation of the shift $W_{1S}$
to the lowest dispersion branch Eq. (\ref{spectrum}) of the exciton. Neglecting
the higher branches requires that $W_{1S} \ll E_B$. To check this requirement,
we use Eq.~(\ref{Gamma}) to present the shift in the form

\begin{equation}
\label{width'}
|W_{1S}|=\Big(\Gamma E_B \Big)^{1/2}.
\end{equation}
The above relation suggests that, when disorder does not destroy the exciton,
i.e. when $\Gamma \ll E_B$, we have $\Gamma \ll W_{1S}\ll E_B$. This, in turn, suggests that
the shift is the dominant effect induced by the disorder. Still, calculating
this shift within the continuum of the ground-state branch is justified.

\section{Higher S-states of an exciton}
Since the width of the 1S state is independent of the form of the wave function, it
is obvious that the width of the 2S state is also equal to $\Gamma$. On the other hand,
the correlator of the effective potential Eq. (\ref{corr2})
acting on the center of mass depends on the form of $\psi_{1S}({\bf r})$. Since
the function $\psi_{1S}({\bf r})$ is more compact than  $\psi_{2S}({\bf r})$,
we expect that the shift for the 2S state is smaller than for the 1S state. Calculation
of this shift is similar to the calculation of
$W_{1S}$. The wave function of the 2S state has the form
\begin{equation}
\label{2S}
\psi_{2S}({\bf r})=
\frac{1}{4(2\pi)^{1/2}a^{3/2}}\left(2-\frac{r}{a}\right)\exp\left(-\frac{r}{2a}  \right).
\end{equation}
Calculation of the Fourier transform of $\psi_{2S}({\bf r})$ is straightforward.
The result reads
\begin{equation}
\label{relation1}
\int d{\bf r}~\psi_{2S}^2\exp(i{\bf qr})=\frac{(1-q^2a^2)(1-2q^2a^2)}{\Big(1+q^2a^2\Big)^4}.
\end{equation}
Substituting this result into the expression Eq. (\ref{shift1S}) for the shift
and performing integration over ${\bf p}'$ using Mathematica, we present the shift
of the 2S state in the form
\begin{equation}
\label{shift2S}
W_{2S}=-D_{2S}\frac{\gamma M^{2}}{\pi^2\hbar^2m_ha},
\end{equation}
where the constant $D_{2S}$ is equal to
\begin{equation}
\label{D2S}
D_{2S}=\frac{77\pi}{1024}.
\end{equation}
This is smaller than the corresponding value for 1S, which is
$D_{1S}=\frac{5\pi}{16}$.
We calculated the value of $D_{3S}$ using Mathematica both for the Fourier transform
and for integration over ${\bf p}'$ and obtained the value
\begin{equation}
\label{D3S}
D_{3S}=\frac{17\pi}{512}.
\end{equation}
For large $n$, the dimensionless shift, $D_{nS}$, can be calculated semiclassically.
The form of the semiclassical wave function is the following
\begin{equation}
\label{semiclassical1}
\psi_{nS}({\bf r})=
A_n\frac{\sin\Big[\left(\frac{2\mu}{\hbar^2} \right)^{1/2}\int\limits_0^r dr'\left( \frac{e^2}{\kappa r'}-|E_n|  \right)^{1/2}                   \Big]}
{r\left( \frac{e^2}{\kappa r'}-|E_n|  \right)^{1/4} },
\end{equation}
where $|E_n|=E_B/n^2$ and $A_n$ is the normalization constant.
In determining $A_n$, the sine square in $\psi_{nS}({\bf r})^2$ can be
replaced by $\frac{1}{2}$. Equally, this replacement is justified
in calculation of the Fourier transform
\begin{align}
\label{Fourier}
&\int d{\bf r}~\psi_{nS}({\bf r})^2\exp{\left(i{\bf qr}\right)}\nonumber\\
&=2\pi\int\limits_0^{r_n}dr
\frac{A_n^2}{\left( \frac{e^2}{\kappa r}-|E_n|  \right)^{1/2}}\Bigg(\frac{2\sin qr}{qr}  \Bigg),
\end{align}
where $r_n=a_Bn^2$ is the radius on $n$-th orbit. The reason why the oscillations
in $\psi_{nS}^2$ can be averaged out is that the characteristic $q$ in Eq.  (\ref{Fourier})
is of the order of $1/r_n$, while $\psi_{nS}$ oscillates $n\gg 1$ times between $r=0$
and $r=r_n$.

Integral Eq. (\ref{Fourier}) can be evaluated explicitly upon
substitution of $r=\frac{1}{2}r_n(1-y)$, after which it
transforms as follows
\begin{equation}
\label{Fourier'}
\int d{\bf r}~\psi_{nS}({\bf r})^2\exp{\left(i{\bf qr}\right)}
=\frac{2\pi A_n^2}{q|E_n|^{1/2}}\int\limits_{-1}^{1}
dy \frac{\sin\Big[\frac{qr_n}{2}(1-y)   \Big]}{(1-y^2)^{1/2}}.
\end{equation}
Notice, that only an even in $y$ term in the numerator of Eq. (\ref{Fourier'})
contributes to the integral. It then reduces to the zero-order Bessel function.
The final result reads
\begin{equation}
\label{Fourier''}
\int d{\bf r}~\psi_{nS}({\bf r})^2\exp{\left(i{\bf qr}\right)}=\frac{\sin\left(\frac{qr_n}{ 2}\right)}{\left(\frac{qr_n}{ 2}\right) }
J_0 \left(\frac{qr_n}{ 2}\right).
\end{equation}
We see that the right-hand-side turns to $1$ in the limit $q\rightarrow 0$, as it
should be. We also see that characteristic $q$ is indeed $\frac{1}{r_n}$. This leads
to the dependence $D_{nS}\propto \frac{1}{r_n} \propto \frac{1}{n^2}$.
With numerical factor specified, we get

\begin{equation}
\label{DnS}
D_{nS}=\frac{2}{n^2}\int\limits_0^{\infty} dx \left(\frac{\sin x}{x}   \right)^2J_0^2(x)=
\frac{1.86}{n^2}.
\end{equation}
Note, however, that increase of $n$ is limited by the condition $R_L \gg r_n$.

\section{2P and 3P-states}
Within the simplest model of the exciton adopted in the
present paper, only the S-states are optically active. Still, P-states
of the exciton can be accessed if the absorption is assisted by the
disorder. Disorder-induced broadening of S and P-states are the same,
but the disorder-induced shifts of the P-states are different.
It is not obvious how the shifts of the levels with the same $n$
are related to each other. To find out, we calculate the shift for
the 2P-states.

We start from the wave function of the 2P-state
\begin{equation}
\label{2Pstate}
\psi_{2P}({\bf r})=\frac{1}{4(2\pi)^{1/2}a^{3/2}}\left(\frac{r}{a}\right)\exp{\left(-\frac{r}{2a}  \right)}\cos \theta,
\end{equation}
where $\theta$ is a polar angle.
Calculation of the Fourier transform is straightforward and yields
\begin{equation}
\label{relation3}
\int d{\bf r}~\psi_{2P}^2\exp(i{\bf qr})=\frac{(1-5q^2a^2}{\Big(1+q^2a^2\Big)^4}.
\end{equation}
The shift is then again calculated using Mathematica. Casting the result
in the form  Eq. (\ref{shift2S}), we find
\begin{equation}
\label{D2P}
D_{2P}=\frac{81\pi}{1024}.
\end{equation}
The fact that $D_{2P}$ exceeds $D_{2S}$ given by Eq. (\ref{D2S})
is our main result. Our calculation shows that the first excited
state, being degenerate in the absence of disorder, is split by
the disorder. Renormalized 2P-level lies below the renormalized 2S-level.
Although the relative splitting is small $\approx 4\%$,
it exceeds parametrically the disorder-induced broadening.

We have also checked that the shifts 3S and 3P levels are close
to each other. Namely, for $D_{3P}$ we obtained the value
\begin{equation}
\label{D3P}
D_{3P}=\frac{199\pi}{6144},
\end{equation}
which is only $2.5\%$ bigger than $D_{3S}$ given by Eq. (\ref{D3S}).


\section{Concluding remarks}

\noindent(i) Behavior of dimensionless shift, $D_{nS}\propto \frac{1}{n^2}$, can be
interpreted as disorder-induced renormalization of the Bohr energy:
\begin{equation}
\label{renormalization}
{\tilde E}_B=E_B\left(1+\frac{1.86\gamma M^2}{\pi^2\hbar^2m_haE_B}   \right).
\end{equation}
To perform an estimate of the correction, we use Eq.~(\ref{width'}).
Within a numerical factor, the relative correction can be cast in the form
$\Big(\Gamma/E_B\Big)^{1/2}$, so that the value of the width and of the Bohr energy can be taken from experiment.\cite{Experiment0} Choosing the values $\Gamma=0.1$mev and
$E_B=100$mev, we get the value $3\%$ for the relative correction.

\vspace{2mm}

\noindent(ii) When calculating the disorder-induced splitting of the 2S and 2P states
we assumed that short-range fluctuations scatter 2S states into 2S states, and the
same fluctuations  scatter 2P states into 2P states. We neglected the contribution
to the shift from $2S \leftrightarrow 2P$ processes. These processes can be taken into
account explicitly if $\psi_{1S}({\bf r})^2$ in Eq. (\ref{shift1S}) is replaced
by the product $\psi_{2S}({\bf r})\psi_{2P}({\bf r})$. The Fourier transform of this
product is given by

\begin{equation}
\label{SP}
\int d{\bf r}\psi_{2S}({\bf r})
\psi_{2P}({\bf r})\exp\left(i{\bf qr}   \right)=\frac{3qa(1-q^2a^2)}{(1+q^2a^2)^4}.
\end{equation}
Calculating the shift from this Fourier transform yields a contribution to $D_{2S}$
equal to $\frac{45\pi}{1024}$, which is comparable to the contribution
$\frac{77\pi}{1024}$ prior. However, it is important to note that
$2S \leftrightarrow 2P$ processes do not contribute to the splitting.
This is because the constant  $\frac{77\pi}{1024}$ adds to $D_{2P}$ as well.
Thus, the $2S \leftrightarrow 2P$ contributions cancel out from the splitting.

More formally, joint renormalizations of S and P exciton branches emerges as diagonal matrix elements of the self-energy matrix
$ \int d{\bf r}\langle {\hat U}{\hat G}_0{\hat U} \rangle$, where ${\hat G}$ is
a bare $2\times 2$ Green function, while $U({\bf r})$ is a perturbation matrix
defined as
\begin{equation}
\label{matrix}
{\hat U}_{ij}=\int d{\bf r}\psi_i({\bf r})\psi_j({\bf r})V_e\Big({\bf R}+\frac{m_h}{M}{\bf r}\Big),
\end{equation}
where subindices $(i,j)$ assume the values S and P.
\vspace{2mm}

\noindent(iii) To put the disorder-induced splitting of 2S and 2P levels of the exciton  into the perspective, it is instructive to invoke
the Lamb shift\cite{Lamb1,Lamb2} in quantum electrodynamics.
In a celebrated experiment\cite{Lamb1} 2S and 2P hydrogen levels
were found to be split by $\approx 4.37$$\mu$eV, a quantity
that can be accounted
for by the electron interaction with vacuum fluctuations.
Theoretically, the shift of a given level is
proportional to $\psi^2(0)$, which is
the probability density to find electron in the state, $\psi({\bf r})$, at the origin\cite{Lamb2}. As a result, the 2P state with $\psi(0)=0$ remains unshifted, while the 2S state lies above the 2P state.
By contrast, for an exciton in a fluctuating field of randomly positioned impurities,
both 2S and 2P levels are shifted, and the 2P level is shifted down more than 2S.

\vspace{2mm}

\noindent(iv) It is well known\cite{Brezin0,Brezin2} that the white-noise disorder leads to the diverging self-energy of the electron which is cut off only by a finite correlation
radius of the disorder. For exciton, the cutoff is provided by a finite Bohr radius, $a$. Our prime observation is that interaction of $2S$ and $2P$ states with the disorder are slightly different.

\vspace{2mm}

\noindent(v) In transition metal dichalcogenides,
see e.g. Ref. \onlinecite{Heinz} and the review Ref. \onlinecite{Glazov'}
a sequence of excited states of exciton in atomically thin samples was identified.
Still, accidental degeneracy of the exciton states in dichalcogenides is missing.
This is because the attraction of an electron and a hole deviates from purely
Coulomb interaction at short distances.\cite{Keldysh}

\section{Acknowledgements}

\vspace{2mm}

The work was supported by the Department of Energy,
Office of Basic Energy Sciences, Grant No.  DE- FG02-
06ER46313.


\vspace{10mm}


\begin{thebibliography}{30}


\bibitem{1952} E. F. Gross and I. A. Karryjew, ``The optical spectrum
of the exciton," Dokl. Akad. Nauk. SSSR {\bf 84}, 471 (1952).
E. F. Gross,  ``Optical spectrum of excitons
in the crystal lattice," Nuovo Cimento Suppl. {\bf 3}, 672
(1956).
\bibitem{1952'}
M. Hayashi1 and K. Katsuki,
``Hydrogen-Like Absorption Spectrum of Cuprous Oxide,"
J. Phys. Soc. Jpn. {\bf 7}, 599 (1952).

\bibitem{1955}
J. H. Apfel and L. N. Hadley,
``Exciton Absorption in Cuprous Oxide,"
Phys. Rev. {\bf 100}, 1689 (1955).

\bibitem{Experiment0}
T. Kazimierczuk, D. Fr{\"o}hlich, S. Scheel, H. Stolz, and M. Bayer,
``Giant Rydberg excitons in the copper oxide Cu$_2$O,"
Nature (London) {\bf 514},
343 (2014).
\bibitem{Glazov}
J. Thewes, J. Heck{\"o}tter, T. Kazimierczuk, M. A$\beta$mann, D. Fr{\"o}hlich, M. Bayer, M. A. Semina, and M. M. Glazov,
``Observation of High Angular Momentum Excitons in Cuprous Oxide,"
Phys. Rev. Lett. {\bf 115}, 027402 (2015).

\bibitem{Experiment1}
F. Schone, S.-O. Kr{\"u}ger, P. Gr{\"u}nwald, H. Stolz, S. Scheel,
M. A$\beta$mann, J. Heckotter, J. Thewes, D. Fr{\"o}hlich, and M. Bayer,
``Deviations of the exciton level spectrum
in Cu$_2$O from the hydrogen series,"
Phys. Rev. B {\bf 93}, 075203 (2016).


\bibitem{microcrystals}
S. Steinhauer, M. A. M. Versteegh, S. Gyger, A. W. Elshaari, B. Kunert, A. Mysyrowicz,  and V. Zwiller,
``Rydberg excitons in Cu$_2$O microcrystals grown on a silicon platform,"
Commun. Mater. {\bf 1}, 11 (2020).

\bibitem{linewidth}
F. Schweiner, J. Main, and G. Wunner,
``Linewidths in excitonic absorption spectra of cuprous oxide,"
Phys. Rev. B {\bf 93}, 085203 (2016).

\bibitem{Chaos}
M. A$\beta$mann,
J. Thewes, D. Fr{\"o}hlich,  and M. Bayer,
``Quantum chaos and breaking of all anti-unitary
symmetries in Rydberg excitons," Nature Materials  {\bf 15}, 741 (2016).

\bibitem{Scaling}
J. Heck{\"o}tter, M. Freitag, D. Fr{\"o}hlich, M. A$\beta$mann,
M. Bayer, M. A. Semina, and M. M. Glazov,
``Scaling laws of Rydberg excitons,"
Phys. Rev. B {\bf 96}, 125142 (2017).

\bibitem{Yakovlev}
P. Rommel, J. Main, A. Farenbruch, D. R. Yakovlev, and M. Bayer,
``Exchange interaction in the yellow exciton series of cuprous oxide,"
Phys. Rev. B {\bf 103}, (2021).

\bibitem{Disorder}
S. O. Kr{\"o}ger, H. Stolz, and S. Scheel,
``Interaction of charged impurities and Rydberg excitons in cuprous oxide,"
Phys. Rev. B {\bf 101}, 235204 (2020).

\bibitem{Theoretical}

``Optical properties of Rydberg excitons and polaritons,"
S. Zieli{\'n}ska-Raczy{\'n}ska, G. Czajkowski, and D. Ziemkiewicz,
Phys. Rev. B {\bf 93}, 075206 (2016).

\bibitem{2D+disorder}
M. Dwedari, S. Brem, M. Feierabend, and E. Malic,
``Disorder-induced broadening of excitonic resonances
 in transition metal dichalcogenides,"
Phys. Rev. Materials {\bf 3}, 074004 (2019).

\bibitem{Alferov}
Zh. I. Alferov, E. L. Portnoi, and A. A. Rogachev,
``Width of the Absorption Edge of Semiconducting Solid Solutions," Fiz. Tech. Poluprovodn. {\bf 2}, 1194 (1968)
[Sov. Phys-Semicond. {\bf 2}, 1001 (1969)].

\bibitem{Baranovskii}
S. D. Baranovskii and A. L. Efros, ``Band Edge Smearing in Solid Solutions,"
 Fiz. Tech. Poluprovodn. {\bf 12}, 2233 (1978)
[Sov. Phys-Semicond. {\bf 12}, 1328 (1978)].

\bibitem{broadening} N. N. Ablyazov, M. E. Raikh, and A. L. Efros,
``Exciton Absorption Linewidth
in Solid Solutions,"
Fiz. Tverd. Tela  {\bf 25}, 253 (1983) [Sov. Phys.-Solid State {\bf 25}, 199
(1983)].

\bibitem{review}
A. L. Efros and M. E. Raikh, ``Effect of Composition Disorder on the Electronic Properties of Semiconducting Mixed Crystals," in
{\em Optical Properties of Mixed Crystals},
edited by R. J. Elliot and I. P. Ipatova (Elsevier, Amsterdam, 1988), p. 133.

\bibitem{Lamb1} W. E. Lamb, Jr. and R. C. Retherford,
``Fine Structure of the Hydrogen Atom by a Microwave Method,"
Phys. Rev. {\bf 72}, 241  (1947).
\bibitem{Lamb2}
H. A. Bethe,
``The Electromagnetic Shift of Energy Levels,"
Phys. Rev. {\bf 72}, 339 (1947).

\bibitem{Brezin0}
E. Brezin and G. Parisi,
``Exponential tail of the electronic density of levels
in a random potential,"
Journ. Phys. C {\bf 13}, L307 (1980).

\bibitem{Brezin2} E. Brezin, D. J. Gross, and D. Itzykson,
``Density of states in the presence of a strong magnetic field
and random impurities," Nucl. Phys. B, {\bf 235}, 24 (1984).

\bibitem{Heinz} A. Chernikov, T. C. Berkelbach,
H. M. Hill, A. Rigosi, Y. Li, O. B. Aslan,
D. R. Reichman, M. S. Hybertsen, and T. F. Heinz,
``Exciton Binding Energy and Nonhydrogenic Rydberg Series
in Monolayer WS$_2$," Phys. Rev. Lett. {\bf 113}, 076802 (2014).

\bibitem{Glazov'}
G. Wang, A. Chernikov, M. M. Glazov, T. F. Heinz,
X. Marie, T. Amand, and B. Urbaszek,
Colloquium: ``Excitons in atomically thin transition metal dichalcogenides,"
Rev. Mod. Phys. {\bf 90}, 021001 (2018).

\bibitem{Keldysh} L. V. Keldysh, ``Coulomb interaction in thin semiconductor
and semimetal films," JETP Lett. {\bf 29}, 658 (1979).

\end{thebibliography}
\end{document}